\documentclass{iau} 
\usepackage{graphicx}

\title[The HMXBs in star-forming galaxies] 
{The High Mass X-ray Binaries in star-forming galaxies}

\author[M. C. Artale, N. Giacobbo, M. Mapelli \& P. Esposito]   
{M. Celeste Artale$^1$, Nicola Giacobbo$^{2,3,4}$, Michela Mapelli$^{1,2,3}$ \& Paolo Esposito$^5$
}

\affiliation{$^{1}$Institut f\"{u}r Astro- und Teilchenphysik, Universit\"{a}t Innsbruck, Technikerstrasse 25/8, 6020 Innsbruck, Austria\\ 
$^{2}$INAF, Osservatorio Astronomico di Padova, vicolo dell'Osservatorio 5, I--35122 Padova, Italy\\
$^{3}$INFN, Milano Bicocca, Piazza della Scienza 3, I--20126, Milano, Italy\\
$^{4}$Dipartimento di Fisica e Astronomia ``G. Galilei'', Universit\`a di Padova, vicolo dell'Osservatorio 3, I-35122, Italy\\
$^{5}$INAF--Istituto di Astrofisica Spaziale e Fisica Cosmica di Milano, via E. Bassini 15, 20133 Milano, Italy \\email: {\tt mcartale@gmail.com}}

\pubyear{2018}
\volume{346} 
\jname{High-mass X-ray binaries: illuminating the passage from massive binaries to merging compact objects}
\editors{A.C. Editor, B.D. Editor \& C.E. Editor, eds.}
\begin{document}

\maketitle

\begin{abstract}
The high mass X-ray binaries (HMXBs) provide an exciting framework to investigate the evolution of massive stars and the processes behind binary evolution.
HMXBs have shown to be good tracers of recent star formation in galaxies and might be important feedback sources at early
stages of the Universe. Furthermore, HMXBs are likely the progenitors of gravitational wave sources (BH--BH or BH--NS binaries that may merge
producing gravitational waves).
In this work, we investigate the nature and properties of HMXB population in star-forming galaxies. 
We combine the results from the population synthesis model MOBSE (\cite[Giaccobo et al. 2018]{Giacobbo2018}) 
together with galaxy catalogs from EAGLE simulation (\cite[Schaye et al. 2015]{Schaye2015}).
Therefore, this method describes the HMXBs within their host galaxies in a self-consistent way.
We compute the X-ray luminosity function (XLF) of HMXBs in star-forming galaxies, showing that this methodology
matches the main features of the observed XLF.  
\end{abstract}

\firstsection 
\section{Introduction}

High Mass X-ray Binaries (HMXBs) are systems composed of a compact object (neutron star NS, or black hole BH) and a massive companion star.
Observational results have shown that HMXBs are good tracers of the star formation rate (SFR) within their host galaxies
(\cite[Grimm et~al. 2003]{Grimm2003}, \cite[Mineo et~al. 2012]{Mineo2012}), 
and might be important heating and ionizing sources in the early Universe (e.g., 
\cite[Justham et~al. 2012]{Justham2012}, \cite[Artale et~al. 2015]{Artale2015}, \cite[Douna et~al. 2018]{Douna2018},
 \cite[Garratt-Simthson et~al. 2018]{Garrat2018}).
From a theoretical point of view, studying the population of HMXBs within galaxies is essential to understand their role in the aforementioned processes and 
the binary evolution.

In particular, the X-ray luminosity function (XLF) is an excellent tracer describing the global population of HMXBs in galaxies.
It can also help to investigate the nature of ultraluminous X-ray sources (\cite[Mapelli et al. 2010]{Mapelli2010}, \cite[Kaaret et al. 2017]{Kaaret2017}). 
Several observational results show that the XLF of HMXBs is described by a power law with a slope
of $\sim 1.6$, and normalization proportional to the SFR (\cite[Grimm et~al. 2003]{Grimm2003}, \cite[Mineo et~al. 2012]{Mineo2012}).

Population synthesis models have proved to be useful to describe the HMXB population of individual galaxies (e.g., \cite[Belczynski et~al. 2004]{Belczynski2004}), 
and to predict the XLF of star-forming galaxies (e.g., \cite[Zuo et al. 2014]{Zuo2014}).
However, they cannot describe the diversity of stellar ages and metallicities within a galaxy in a self-consistent way.

In order to properly model star formation and metallicity evolution in galaxies, population synthesis simulations must
be coupled with galaxy catalogs from galaxy formation models. 
Such galaxy catalogs can be obtained either from semianalytic models (\cite[Fragos et~al. 2013]{Fragos2013}), or from hydrodynamical cosmological simulations
(Mapelli et al. 2017, 2018a, 2018b, Artale et al. in preparation).

In this work, we study the XLF of star-forming galaxies combining the galaxy catalogs of the hydrodynamical cosmological simulation {\sc eagle} 
(\cite[Schaye et al. 2015]{Schaye2015}) with the results from the population synthesis model {\sc mobse} (\cite[Giacobbo et al. 2018a]{Giacobbo2018}). 
In Section~\ref{sec:model} we present the methodology. We discuss our findings in Section~\ref{sec:results}.

\section{Simulations and methodology}\label{sec:model}

\begin{figure}[b]
\centering
 \includegraphics[width=2.4in]{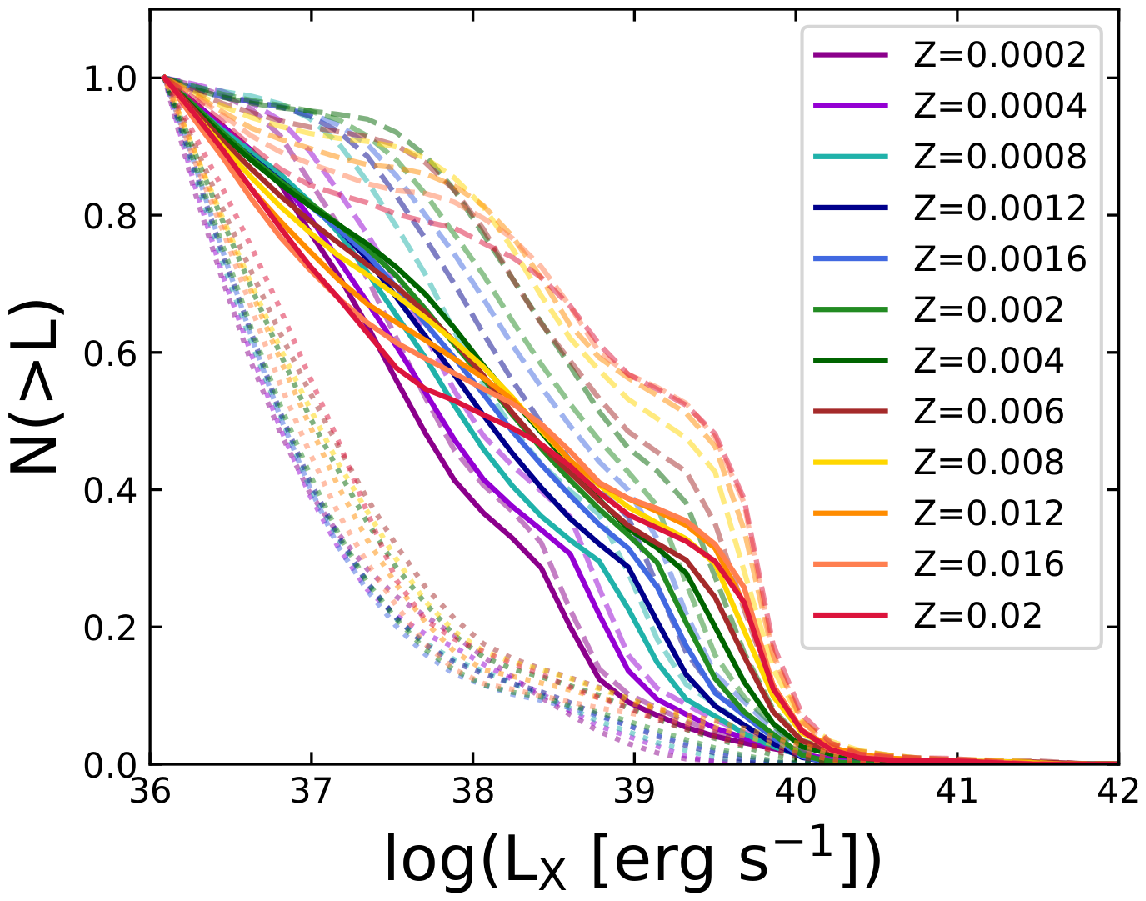}
 \includegraphics[width=2.45in]{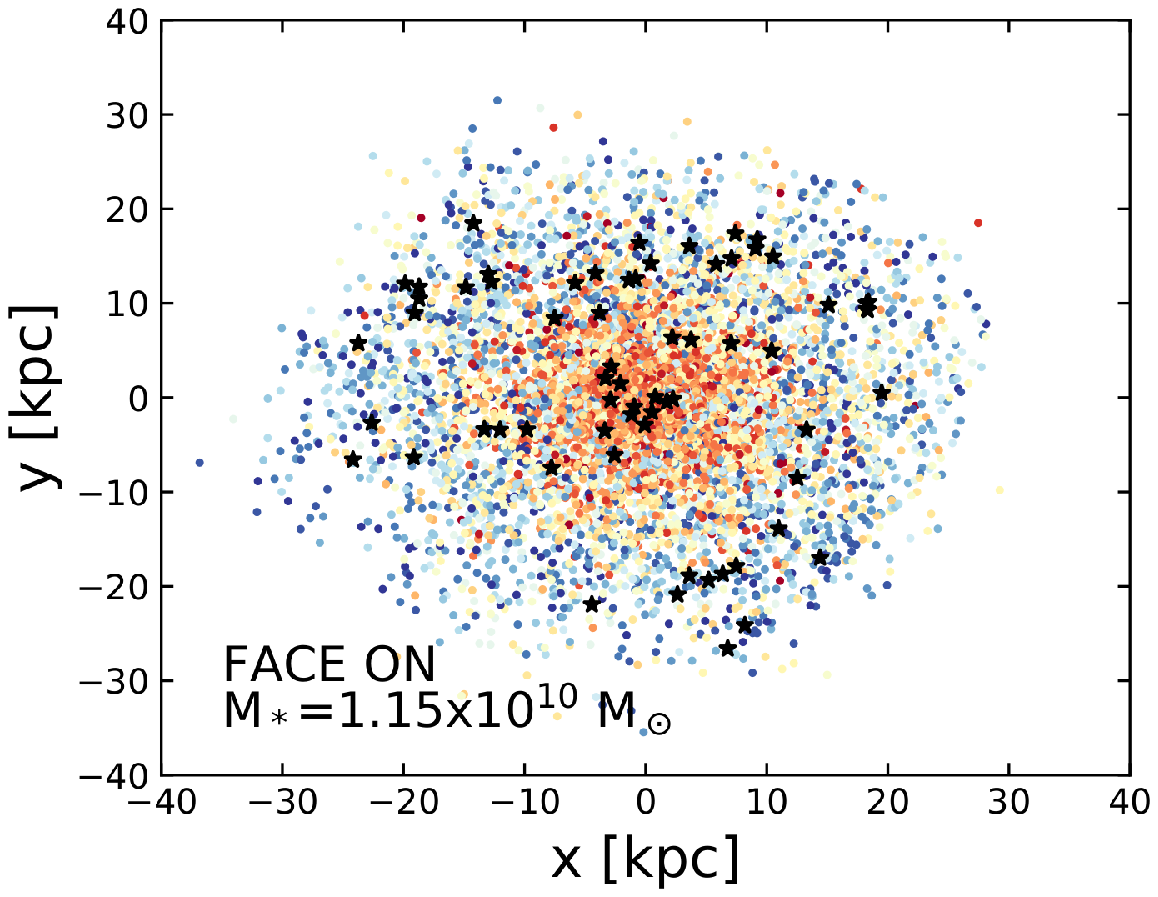} 
 \caption{\textit{Left panel:} Cumulative number of HMXBs normalized by the total number of sources, obtained with {\sc mobse} for the 12 adopted metallicities (filled lines).
 We also show the contribution from SW-HMXBs (dotted lines) and RLO-HMXBs systems (dashed lines). 
 \textit{Right panel:} Spatial distribution of the stellar particles within one of the galaxies in the {\sc eagle} catalog, showed face on. 
 The color code represents the ages of the stellar particles (the oldest and the youngest populations are marked in red  and in blue, respectively).
 Stellar particles that were formed in the last 100 Myr
are shown by black stars (these particles are populated with HMXBs, see the details in Section~\ref{sec:model}).}
 \label{fig1}
\end{figure}

{\sc mobse} (\cite[Giacobbo et al. 2018a]{Giacobbo2018}) is an upgraded version of 
{\sc bse} code (\cite[Hurley et al. 2002]{Hurley2002}).
The code includes new stellar winds prescription (\cite[Vink et al. 2001,2005]{Vink2001,Vink2005}, \cite[Chen et al. 2005]{Chen2015}), 
electron-capture SNe (\cite[Giaccobo \& Mapelli 2018b]{Giacobbo2018b}), 
core-collapse SNe (\cite[Fryer et al. 2012]{Fryer2012}), pulsational pair-instability and pair-instability SNe (\cite[Spera \& Mapelli 2017]{Spera2017}).
{\sc mobse} reproduces successfully the masses and merger rates of compact objects (\cite[Giacobbo et al. 2018b, 2018c]{Giacobbo2018b,Giacobbo2018})
inferred by the LIGO-Virgo collaboration.
In this work we adopt the simulation set referred to as $\alpha{}1$ in \cite[Giacobbo et al. (2018b)]{Giacobbo2018b}.
In this model, the common envelope parameter is set to  $\alpha{}=1$.
This set is composed of 12 sub-sets at different metallicities Z = 0.0002, 0.0004, 0.0008,
0.0012, 0.0016, 0.002, 0.004, 0.006, 0.008, 0.012, 0.016 and 0.02.
Each sub-set contains $10^6$ binaries, hence the total number of binaries is $1.2\times10^{7}$.
From the population synthesis model, we identify the HMXB sources undergoing stable mass transfer via Roche lobe overflow (RLO--HMXB),
and those accreting the wind from the companion star (SW--HMXB).
The X-ray luminosity of each HMXB in the catalog is computed as $L_{\rm X} = \eta \frac{G \dot{M}_{\rm acc} M_{\rm co}}{R_{\rm co}}$,
where $\dot{M}_{\rm acc}$, $M_{\rm co}$, and $R_{\rm co}$ are the accretion rate, the mass, and the radius of the compact object, respectively. 
The parameter $\eta$ is the efficiency in converting gravitational binding energy to radiation associated with accretion. 
We adopt that $\eta = 0.1$ for BH and NS for simplicity since BH--HMXB sources are the dominant population (see ahead in the text). 
On the left panel of Figure~\ref{fig1}, we show the cumulative distribution of HMXBs normalized by the total number of sources for each metallicity sub-set.
We also split the contribution of SW and RLO systems in each subsample.

The {\sc eagle} simulation suite is a set of cosmological hydrodynamical simulations
with different resolution levels and box sizes, run using an updated version of {\sc gadget-3} code.
It adopts the $\Lambda$CDM cosmology with cosmological parameters $\Omega_{\rm m} = 0.2588$, $\Omega_{\Lambda} = 0.693$,
$\Omega_{\rm b} = 0.0482$, and ${\rm H_{0}} = 100~h~{\rm km~s^{-1} Mpc^{-1}}$ with $h = 0.6777$ (\cite[Planck Collaboration 2014]{Planck2014}).
The simulation includes subgrid models accounting for star formation, UV/X-ray ionizing background, radiative cooling and heating, 
stellar evolution, chemical enrichment, AGB stars and SNe feedback, and supermassive black hole feedback.
In this work we use the simulated box named as L0100N1504. This run represents a periodic box of 100~Mpc side, which initially
contains $1504^3$ gas and dark matter particles with masses of
$m_{\rm gas} = 1.23\times10^{6} h^{-1} {\rm M_{\odot}}$ and $m_{\rm dm}=6.57\times10^{6} h^{-1} {\rm M_{\odot}}$.
Since this work is focused on the analysis of star forming galaxies in the local universe, we use
the galaxy catalog at $z=0$. We select a subsample of galaxies
with specific star formation rate of $sSFR > 10^{-10} {\rm M_{\odot} yr^{-1}}$ and stellar masses in the range of
${\rm M}_{*} = 10^{8} - 5\times10^{10}~{\rm M_{\odot}}$. 
The number of galaxies fulfilling this condition is 2596.

For each simulated galaxy in the subsample, we identify the youngest stellar particles with ages below to 100~Myr.
Since the progenitors of HMXBs are systems composed of two massive stars, these sources are directly connected with the star-forming regions within the galaxies.
The assumption to select the youngest stellar particles ($< 100$~Myr) accounts for this fact. 
In Figure~\ref{fig1} right panel, we show the spatial distribution of the stellar particles
in one of the galaxies of the sub-sample. 
We find that the stellar particles with age $<100$~Myr (indicated by black stars in Figure~\ref{fig1}) are mainly in the outskirts of this galaxy,
and a few of them are located close to the central region.

Hence, for each galaxy in the subsample, we identify the stellar particles that fulfil the age condition, and according to 
the metallicity of each particle $Z_*$, we compute the number of HMXBs as \cite[Mapelli et al. (2010)]{Mapelli2010},

\begin{equation}\label{eq:number}
 N_{\rm HMXB} = \frac{N_{\rm HMXB}^{\rm MOBSE}(Z_*)}{m^{\rm MOBSE}(Z_*)} m_{*}^{\rm EAGLE} f_{\rm corr} f_{\rm bin},
\end{equation}

\noindent where  $m_{*}^{\rm EAGLE}$ is the mass of the stellar particle, ${\rm N}_{\rm HMXB}^{\rm MOBSE}(Z_*)$ is the total number of HMXBs 
in the {\sc mobse} catalog with the metallicity closer to $Z_*$, and $m^{\rm MOBSE}(Z_*)$ is the total mass of the {\sc mobse} catalog at the selected metallicity.
The parameter $f_{\rm corr}=0.285$ accounts for the fact that we simulate only massive stars ($\ge{}5$ M$_\odot$), while $f_{\rm bin}=0.5$ is the assumed binary fraction.
Hence,  we randomly select a number $N_{\rm HMXB}$ of HMXBs from the sub-set of {\sc mobse} with the metallicity closer to $Z_\ast$ and we assign them to that star particle.

Our model also accounts for transient and persistent sources and includes prescriptions for Be--HMXB systems.
Following \cite[Zuo et al. (2014)]{Zuo2014}, we assume that Be--HMXBs
are wind-fed systems composed of an NS and a massive companion star, with orbital periods in the range of 10-300~d.
We assume that only 25\% of these systems are Be--HMXBs.  
We identify transient sources through the thermal disk instability model, where binaries with accretion rates below a critical value
$\dot{M}_{\rm crit}$ are considered transient sources (see \cite[Frank, King and Raine 2002]{Frank2002}, eq. 5.105 and 5.106, p. 133).
We note here that the assumptions made for transient sources are based on models for low-mass X-ray binary sources.
Transient sources are in a quiescent state most of the time. Hence, we assume that its duty cycle is 1\%.
We also adopt a bolometric correction following \cite[Fragos et al. (2013)]{Fragos2013}.

We compute the error bars assuming a Poisson distribution for the X-ray luminosities within the galaxies. 
We split the mean XLF of the star-forming galaxies according to the compact object (BH--HMXB and NS--HMXB)
and the accretion process (RLO--HMXB, SW--HMXB, Be--HMXB) of the sources.

Using this method, for each simulated galaxy we obtain a population of HMXBs that accounts for the variability of the sources
and the different emission mechanisms.
In this work, we focus on studying the XLF of star-forming galaxies and compare our findings with
the observational results of \cite[Mineo et al. (2012)]{Mineo2012}.

\section{Results and future work}\label{sec:results}

\begin{figure}[b]
\begin{center}
 \includegraphics[width=3.4in]{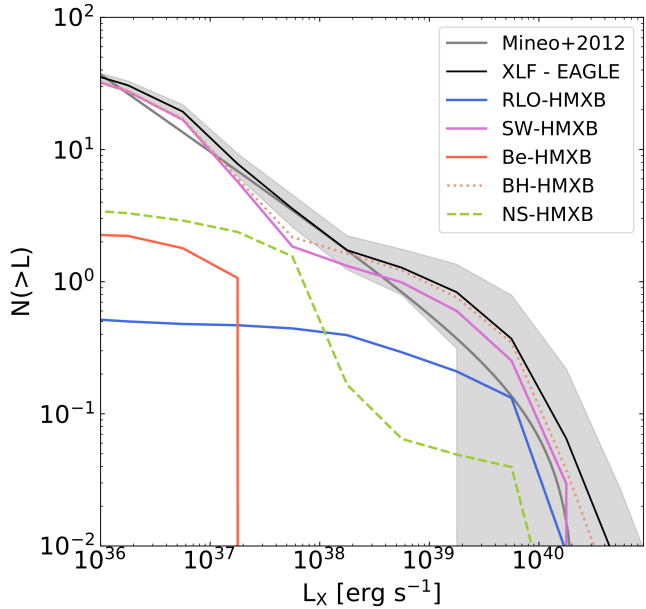} 
 \caption{Mean X-ray luminosity function (XLF) normalized by the galaxy star formation rate.
Black line: Mean XLF obtained by stacking together the XLFs of the star-forming galaxies in the EAGLE simulation at $z=0$.
Shaded grey area: Poissonian uncertainty on the mean XLF from the EAGLE galaxies.
Solid blue (magenta) line: contribution to the simulated XLF by RLO--HMXBs (SW--HMXBs).
Dotted red (dashed green) line: BH--HMXBs (NS--HMXBs).
Solid red line: Be-HMXBs. 
Grey line: observed XLF by \cite[Mineo et al. (2012)]{Mineo2012}.}
 \label{fig2}
\end{center}
\end{figure}

Figure~\ref{fig2} shows the mean XLF obtained by stacking together the XLFs of the star-forming galaxies.
Our results show that the mean XLF of the simulated galaxies is in fair agreement
with the observed XLF by  \cite[Mineo et al. (2012, grey line)]{Mineo2012}.

In our model, BH--HMXBs are more numerous and generally brighter than NS--HMXBs.
BH systems are expected to be more luminous than NS--HMXBs (\cite[Kaaret \etal\ 2017]{Kaaret2017}).
However, observational results indicate that persistent NS--HMXBs are more numerous than persistent BH--HMXBs in the Milky Way (e.g., \cite[Lutovinov et al. 2013]{Lutovinov2013}).

Nonetheless, \cite[Vulic et al. (2018)]{Vulic2018}
show that galaxies with sSFR $>10^{-10}~{\rm yr}^{-1}$ have a higher number of BH-HMXBs than NS-HMXBs due to recent star formation episodes. 
Moreover, the fraction of BH-HMXBs and NS-HMXBs from the population synthesis model output shows that BH--HMXBs are more numerous in the simulated set. 
This is explained since strong interaction in binary systems tends to form more BH than NS due to mass transfer.

When we compare the HMXB according to the accretion process, we find that the
RLO--HMXBs contribute only with high X-ray luminosity sources, while
while SW-HMXBs dominate in all the X-ray luminosity range.

Several parameters in the population synthesis model (e.g. supernova kicks)
might play a fundamental role in shaping the population of HMXBs.
In a forthcoming work, we will investigate in detail the impact of some key population-synthesis parameters on the demography of HMXBs.

​\section*{Ackwnoledgements}
MCA acknowledges financial support from the Austrian National Science Foundation through FWF stand-alone grant P31154-N27.


\begin{thebibliography}{}

\bibitem[Artale \etal\ (2015)]{Artale2015}
{Artale, M.~C., Tissera, P.~B., Pellizza, L.~J} 2015,
\textit{MNRAS}, 448, 3071

\bibitem[Belczynski \etal\ (2004)]{Belczynski2004}
{Belczynski, K., Kalogera, V., Zezas, A., Fabbiano, G.} 2004,
\textit{APJL}, 601, 147

\bibitem[Chen \etal\ (2015)]{Chen2015}
{Chen, Y., Bressan, A., Girardi, L., Marigo, P., Kong, X., Lanza, A.} 2015,
\textit{MNRAS}, 452, 1068

\bibitem[Douna \etal\ (2018)]{Douna2018}
{Douna, V.~M., Pellizza, L.~J., Laurent, P., Mirabel, I.~F.} 2018
\textit{MNRAS}, 474, 3488

\bibitem[Fragos \etal\ (2013)]{Fragos2013}
{Fragos, T., Lehmer, B., Tremmel, M., Tzanavaris, P., Basu-Zych, A., Belczynski, K., Hornschemeier, A., Jenkins, L., Kalogera, V., Ptak, A. , Zezas, A.} 2013,
\textit{ApJ}, 764, 41

\bibitem[Frank \etal\ (2002)]{Frank2002}
{Frank, J., King, A., Raine, D.~J.} 2002,
\textit{Accretion Power in Astrophysics}
ISBN 0521620538.~Cambridge, UK. Cambridge University Press

\bibitem[Fryer \etal\ (2012)]{Fryer2012}
{Fryer, C.~L., Belczynski, K., Wiktorowicz, G., Dominik, M., Kalogera, V., Holz, D.~E.} 2012, 
\textit{ApJ}, 749, 91

\bibitem[Garratt-Simthson \etal\ (2018)]{Garrat2018}
{Garratt-Smithson, L., Wynn, G.~A., Power, C., Nixon, C.~J.} 2018,
\textit{MNRAS}, 480, 2985

\bibitem[Giacobbo \& Mapelli (2018)]{Giacobbo2018}
{Giacobbo, N., Mapelli, M., Spera, M.} 2018a,
\textit{MNRAS}, 474, 2959 

\bibitem[Giacobbo \& Mapelli (2018)]{Giacobbo2018b}
{Giacobbo, N., Mapelli M.} 2018b,
preprint (arXiv: 1805.11100)

\bibitem[Giacobbo \etal\ (2018)]{Giacobbo2018c}
{Giacobbo, N., Mapelli, M.} 2018c,
\textit{MNRAS}, 480, 2011

\bibitem[Grimm \etal\ (2003)]{Grimm2003}
{Grimm, H.-J., Gilfanov, M., Sunyaev, R.} 2003,
\textit{MNRAS}, 339, 793

\bibitem[Hurley \etal\ (2002)]{Hurley2002}
{Hurley J.~R., Tout C.~A., Pols O. R.} 2002
\textit{MNRAS} 329, 897

\bibitem[Justham \& Schawinski (2012))]{Justham2012}
{Justham, S. \& Schawinski, K.} 2012,
\textit{MNRAS}, 423, 1641

\bibitem[Kaaret \etal\ (2017)]{Kaaret2017}
{Kaaret, P., Feng, H., Roberts, T.~P.} 2017,
\textit{ARAA}, 55, 303

\bibitem[Lutovinov \etal\ (2013)]{Lutovinov2013}
{Lutovinov, A.~A., Revnivtsev, M.~G., Tsygankov, S.~S., Krivonos, R.~A.} 2013,
\textit{MNRAS}, 431, 327

\bibitem[Mapelli \etal\ (2010)]{Mapelli2010} 
{Mapelli, M., Ripamonti, E., Zampieri, L., Colpi, M., Bressan, A.} 2010,
\textit{MNRAS}, 408, 234

\bibitem[Mapelli \etal\ (2017)]{Mapelli2017}  
{Mapelli, M., Giacobbo, N., Ripamonti, E., Spera, M.} 2017,
\textit{MNRAS}, 472, 2422 

\bibitem[Mapelli \& Giacobbo (2018a)]{Mapelli2018a}
{Mapelli, M., \& Giacobbo, N.} 2018a,
\textit{MNRAS}, 479, 4391 

\bibitem[Mapelli \etal\ (2018b)]{Mapelli2018b} 
{Mapelli, M., Giacobbo, N., Toffano, M., et~al.} 2018b, 
arXiv:1809.03521 

\bibitem[Mineo \etal\ (2012)]{Mineo2012}
{Mineo, S., Gilfanov, M., Sunyaev, R.} 2012,
\textit{MNRAS}, 419, 2095 

\bibitem[Planck Collaboration (2014)]{Planck2014}
{Planck Collaboration} 2014,
\textit{A\&A}, 571, 16

\bibitem[Spera \etal\ 2017]{Spera2017}
{Spera, M., Mapelli, M.} 2017,
\textit{MNRAS}, 470, 4739

\bibitem[Schaye \etal\ (2015)]{Schaye2015}
{Schaye, J., Crain, R.~A., Bower, R.~G., Furlong, M., Schaller, M., Theuns, T., Dalla Vecchia, C., Frenk, C.~S., McCarthy, I.~G.,
 Helly, J.~C., Jenkins, A., Rosas-Guevara, Y.~M., et al.} 2015,
 \textit{MNRAS}, 446, 521 
 
\bibitem[Vink \etal\ (2001)]{Vink2001}
{Vink, J.~S., de Koter, A., Lamers, H.~J.~G.~L.~M.} 2001,
\textit{A\&A}, 369, 574

\bibitem[Vink \etal\ (2005)]{Vink2005}
{Vink, J.~S., de Koter, A.} 2005,
\textit{A\&A}, 442, 587

\bibitem[Vulic \etal\ (2018)]{Vulic2018}
{Vulic, N., Hornschemeier, A.~E., Wik, D.~R., Yukita, M., Zezas, A., Ptak, A.~F., Lehmer, B.~D., et al.} 2018,
\textit{ApJ}, 864, 150

\bibitem[Zuo \etal\ (2014)]{Zuo2014}
{Zuo, Z.-Y., Li, X.-D., Gu, Q.-S.} 2014,
\textit{MNRAS}, 437, 1187

\end{thebibliography}
\end{document}